\documentclass[pdflatex,sn-mathphys-num]{sn-jnl}%

\usepackage{graphicx}
\usepackage{multirow}
\usepackage{amsmath,amssymb,amsfonts}
\usepackage{amsthm}
\usepackage{mathrsfs}
\usepackage[title]{appendix}
\usepackage{xcolor}
\usepackage{textcomp}
\usepackage{manyfoot}
\usepackage{subcaption}
\usepackage{algorithm}
\usepackage{algorithmicx}
\usepackage{algpseudocode}
\usepackage{csquotes}
\usepackage{listings}
\usepackage{xspace}
\usepackage{tabularx}
\usepackage{booktabs}
\usepackage{tikz}
\usepackage{quantikz}
\usepackage{comment}
\usepackage{braket}

\newcommand{\eg}{\emph{e.g.}\xspace}
\newcommand{\ie}{\emph{i.e.}\xspace}
\newcommand{\etal}{\emph{et al.}\xspace}
\newcommand{\repro}{\href{https://github.com/lfd/DSE.git}{reproduction package}\xspace}
\newcommand{\zenodo}{\href{https://zenodo.org/records/15557369}{Zenodo}\xspace}

\begin{document}

\title[Stacking the Odds]{Stacking the Odds: Full-Stack Quantum System Design Space Exploration}


\author*[1,4]{\fnm{Hila} \sur{Safi}}\email{hila.safi@siemens.com}
\equalcont{These authors contributed equally to this work.}

\author*[2, 3]{\fnm{Medina} \sur{Bandic}}\email{m.bandic@tudelft.nl}
\equalcont{These authors contributed equally to this work.}
\author[4]{\fnm{Christoph} \sur{Niedermeier}}\email{christoph.niedermeier@siemens.com}
\author[5]{\fnm{Carmen} \sur{G. Almudever}}\email{cargara2@disca.upv.es}
\author[2, 3]{\fnm{Sebastian} \sur{Feld}}\email{s.feld@tudelft.nl}
\author[1, 4]{\fnm{Wolfgang} \sur{Mauerer}}\email{wolfgang.mauerer@othr.de}

\affil[1]{\orgdiv{Laboratory for Digitalisation},\orgname{Technical University
of Applied Sciences Regensburg}, \orgaddress{\street{Seybothstr.~2}, \city{Regensburg},
\postcode{93053}, \state{Bavaria}, \country{Germany}}}

\affil[2]{\orgdiv{Quantum \& Computer Engineering}, \orgname{Delft University
of Technology}, \orgaddress{\street{Mekelweg 5}, \city{Delft}, \postcode{2628 CD},
\country{Netherlands}}}

\affil[3]{\orgname{QuTech}, \orgaddress{\street{Lorentzweg 1}, \city{Delft},
\postcode{2628 CJ}, \state{Delft}, \country{Netherlands}}}

\affil[4]{\orgdiv{Foundational Technology}, \orgname{Siemens AG},
\orgaddress{\street{Friedrich-Ludwig-Bauer-Str.~3}, \city{Garching bei München},
\postcode{85748}, \state{Bavaria}, \country{Germany}}}

\affil[5]{\orgdiv{Computer Engineering}, \orgname{Universitat Politècnica de València},
\orgaddress{\street{Camí de Vera}, \city{València}, \postcode{46022}, \country{Spain}}}


\abstract{Design space exploration (DSE) plays an important role in optimising quantum circuit execution
by systematically evaluating different configurations of compilation strategies and hardware 
settings. In this paper, we conduct a comprehensive investigation into the impact of various
layout methods, qubit routing techniques, and optimisation levels, as well as device-specific properties
such as different variants and strengths of noise and imperfections, the topological structure of qubits, connectivity densities, and back-end sizes.
By spanning through these dimensions, we aim to understand the interplay between compilation choices and 
hardware characteristics. A key question driving our exploration is whether the optimal selection of device
parameters,  mapping techniques, comprising of initial layout strategies and routing heuristics can mitigate
device induced errors beyond standard error mitigation approaches. Our results show that carefully selecting software strategies
(\eg, mapping and routing algorithms) and tailoring hardware characteristics (such as minimising noise and leveraging topology and connectivity density) significantly improve
the fidelity of circuit execution outcomes, and thus the expected correctness or success probability of the computational result. We provide estimates based on key metrics such as circuit depth, 
gate count and expected fidelity.
Our results highlight the importance of 
hardware–software co-design, particularly as quantum systems scale to larger dimensions,
and along the way towards fully error corrected quantum systems:
Our study is based on computationally noisy simulations,
but considers various implementations of
quantum error correction (QEC) using the same approach as for other algorithms.
The observed sensitivity of circuit fidelity
to noise and connectivity suggests that co-design principles will be equally
critical when integrating QEC in future systems. Our exploration provides practical
guidelines for co-optimising physical mapping, qubit routing, and hardware
configurations in realistic quantum computing scenarios.}


\maketitle
\section{Introduction}\label{sec:Introduction}
As quantum computing moves closer to practical application, the 
performance bottlenecks are no longer defined solely by hardware limitations
or algorithmic efficiency, but increasingly by how well both align~\cite{on-the-co-design}.
Current devices suffer from limited qubit counts, restricted connectivity, and different
types of noise, all of which degrade circuit performance and solution quality.
Just as classical computing has benefited from decades of hardware-software co-design, 
quantum computing demands a similar approach~\cite{tomesh2021quantum, on-the-co-design, wintersperger:22:codes}.
In this work, we take a layered perspective on quantum circuit compilation (transpilation)\cite{qiskit_transpiler},
examining how choices made across the hardware-software stack collectively shape 
performance and impact fidelity and resource efficiency. We apply a Design Space Exploration (DSE) approach to systematically
evaluate these strategies. DSE enables the identification of optimal
trade-offs by balancing the exploration of novel design configurations
with exploitation of known high-performing strategies~\cite{bandic2020structured}.
As illustrated in Table~\ref{tab:softwarestack}, we span
five layers---from the initial circuit setup, through qubit mapping and routing, down
to noise modelling and hardware topology. At each layer, we systematically explore the 
design space using a set of representative benchmark circuits covering algorithmic,
variational, arithmetic, and simulation-based circuits to enable a meaningful
evaluation.
These include well-known algorithms, for instance Quantum Fourier Transform (QFT), Shor's
algorithm, or VQE~\cite{bandic2023interaction,qbench,article_profiling_bandic}. To reflect the ongoing research
toward fault-tolerant quantum computing, we extend our evaluations to include
quantum error correction codes---thus aligning near-term compilation strategies
with future fault-tolerant demands. We begin our investigation with device-level parameters,
including back-end size, qubit connectivity, topologies and different noise variants---
with a focus on crosstalk. Crosstalk is a dominant and often underestimated source
of correlated error~\cite{sarovar2020detecting}. All hardware aspects influence gate scheduling, circuit depth, error rates, and routing complexity.
Building on this, we examine the mapping layer, and evaluate how different initial layout strategies,
qubit routing techniques, and optimisation levels (as implemented in Qiskit ~\cite{qiskit_transpiler}) impact
circuit depth and fidelity. On top of these, we incorporate multiple
additional optimisations (\eg gate simplification techniques).
Our results confirm that optimal circuit compilation is not only
back-end-dependent in terms of architecture, but also strongly influenced
by hardware-specific noise characteristics such as decoherence
and crosstalk. Moreover,
we show that informed decisions on quantum circuit mapping can reduce 
the effect of noise, achieving improvements in addition to conventional error mitigation techniques like Zero Noise Extrapolation~\cite{Giurgica_Tiron_2020}. We present a layered DSE framework for quantum circuit
optimisation, highlight the critical impact of connectivity and noise variations,
particularly crosstalk, and show that co-optimised mapping settings improve
circuit fidelity and resource efficiency.
By benchmarking multiple quantum algorithms across this multi-layered stack, our work
offers actionable guidelines for future quantum system design. Taken together,
our contributions demonstrate the necessity for treating quantum circuit
compilation as a full-stack optimisation problem. Rather that tuning
individual parameters in isolation, we advocate for 
a holistic, hardware-aware approach. This layered perspective
enables more informed design choices, ultimately leading to 
higher-fidelity executions and more efficient use of limited
quantum resources.
The paper is augmented by a~\repro~\cite{mauerer:22:q-saner},
(link in PDF) that also contains the full set of benchmarks used in each experiment.
For long-term reproducibility and archival, we also provide a snapshot
via~\zenodo.

\begin{table}[htbp]
    \caption{Full-stack design space (parameters and techniques) explored in our work.}\label{tab:softwarestack}
    \begin{tabularx}{\linewidth}{XX}
  \toprule
  Layers & Examples \\\midrule
Algorithmic Design \& Circuit Setup & Problem formulation, QAOA layers\\
  Logical Circuit Optimisation & Gate count reduction, optimise Clifford Gates\\
  Qubit Mapping \& Routing & Qubit allocation, routing, swap gate insertion\\
  Noise Modelling \& Hardware Simulation & Depolarisation, crosstalk\\
   Topology & Heavy-hex, Sycamore\\
\bottomrule
  \end{tabularx}
\end{table}

\section{Context, Foundation \& Prior Work}
The execution of quantum circuits on current NISQ (Noisy Intermediate-Scale Quantum) 
devices presents significant challenges due to hardware limitations, error-prone operations,
and restricted qubit connectivity. Addressing these constraints requires a full-stack quantum
computing approach, where both the quantum hardware and software stack are co-designed to
enhance performance and scalability. Safi~\etal~\cite{Safi_10234303} demonstrate
that co-designing quantum processing units (QPUs) for specific applications can significantly
improve execution performance, even on relatively simple architectures. 
Complementary to hardware-focused efforts, recent work~\cite{thelen:designautomation}
introduces software-layer co-design methodology that automates the
selection of quantum-classical algorithms and their parameters based on
non-functional requirements, thereby supporting scalable
and application-aware quantum software development.
Bandic~\etal~\cite{bandic2022full} argue that end-to-end integration---from qubit
control hardware to compilers---is essential for effective noise mitigation
and progress toward fault tolerance. Such studies highlight the importance of co-optimisation
across the stack to align hardware capabilities with software demands.
At the device level, critical parameters such as back-end size, qubit count, native gate set,
connectivity, and noise characteristics profoundly influence circuit execution. Optimising
these properties has been the focus of several works. For instance, Murali et al.~\cite{murali2019noise}
proposed a noise-adaptive strategy that selects qubits based on individual error rates to increase
execution fidelity. Li et al.~\cite{li2020towards} demonstrated that topology-aware design
decisions---such as application-specific qubit placement, customised coupling graph 
topologies, and routing-aware compilation---can reduce the resource overhead needed for realistic workloads. The impact of
these parameters extends into the software stack, where hardware-aware compilers leverage
device knowledge to guide circuit optimisation~\cite{murali2019full}.
Furthermore, the MQT Predictor framework by Quetschlich~\etal~\cite{10.1145/3673241_quetschlich}
introduces an automated approach
to selecting quantum devices and optimising compilation flows, also
demonstrating significant performance gains.
Although such techniques
are relevant for near-term devices, circuit optimisation
remains equally critical for fault-tolerant devices~\cite{Amy_2013}. Our 
work explores the intersection of hardware-level parameters and compilation
strategies to identify and evaluate the optimal balance, that maximised circuit fidelity
with realistic architectural constraints.
\subsection{NISQ Architecture vs. FTQC}
NISQ era, coined by Preskill~\cite{preskill2018quantum},
describes today’s quantum devices with
50 to 1000 qubits that suffer from noise and decoherence. Although not
fully error-corrected, an intensive discussion has unfolded on the
computational advantage of NISQ machines through variational methods~\cite{cerezo2021variational} or the QAOA  class of algorithms~\cite{farhi2014quantumapproximateoptimizationalgorithm, thelen:24:noisy-qaoa}.
Despite being unsuitable
for large-scale fault-tolerant tasks, error mitigation techniques can improve
their practical utility~\cite{Temme_2017}. Fault-tolerant quantum computing (FTQC)~\cite{548464_shor} overcomes these limitations using quantum error correction (QEC),
e.g., surface codes~\cite{Fowler_2012}, which encode logical qubits across many
physical ones. FTQC demands high qubit fidelity and scale. Bridging NISQ and
FTQC will require hybrid strategies---combining classical and quantum processing---
, resilient algorithms, improved connectivity,
and crosstalk-tolerant QEC codes~\cite{Roffe_2019, Safi_10234303}.

\subsection{DSE for Hardware-Software Co-Design}
In the pursuit of practical quantum computing applications, hardware-software
co-design (HW-SW co-design) is a promising strategy, particularly in light of the current limitations
of quantum hardware, such as restricted number of qubits and high error rates---especially
in \emph{two-qubit} gate operations~\cite{Safi_10234303,bandic2022full,
murali2019full,bridge1,bridge2}.
These constraints are a significant challenge for executing
complex quantum algorithms in a reliable way.
HW-SW co-design involves the collaborative development of quantum hardware and software, 
ensuring that both components are optimised
in tandem to enhance overall system performance. By aligning the design of quantum
algorithms with the specific characteristics of QPUs,
and vice versa, co-design facilitates more efficient and effective quantum computations~\cite{wintersperger:22:codes,thelen:24:noisy-qaoa}.
In our research, we employ design space exploration as a systematic approach to 
optimise quantum circuit compilation and assess the hardware characteristics necessary
for efficient execution of benchmark problems~\cite{bandic2020structured}.
DSE, in general, enables structured evaluation of architectural and algorithmic design 
choices---such as in our case qubit mappings, circuit decompositions, and noise-aware 
strategies for error reduction---with the goal of enhancing performance,
scalability and resource efficiency. Within this framework, quantum circuit
compilation refers to the process of transforming a high-level circuit into a low-level,
hardware-executable form, involving qubit mapping
and operations scheduling.
Among the key compiler-level challenges, quantum circuit mapping plays
a fundamental role in enabling full-stack quantum computing~\cite{li2019tackling,almudever2020realizing}.
Efficient mapping is supposed to reduce gate overhead
and latency, but becomes increasingly challenging as 
qubit count grows. To optimise these metrics, various approaches---ranging from
heuristics and brute-force methods to graph-based, dynamic programming, and machine learning
techniques---have been developed~\cite{pozzi2020using,
zulehner2018efficient,wagner2023improving,bandic2023mappingqubo,
ovide2023mapping,OnQubitMappingProblem}. We systematically
analyse critical metrics such as circuit depth and gate counts,
as well as the impact of various noise sources---particularly crosstalk---on overall
circuit performance. We observed performance variations stem from key hardware properties,
including topology, layout strategies, qubit routing techniques, and optimisation levels.
Table~\ref{tab:softwarestack} illustrates the key hardware-aware design choices involved
in our benchmarking approach, spanning logical circuit optimisation, qubit mapping, noise modelling,
and hardware back-end selection. Although we do not
modify the circuit design in our study, we incorporate a broad range of circuits, including error-correcting ones,
to assess their performance across different compilation and execution strategies. This ensures
that our findings remain relevant not only for current quantum devices but also as a preparatory
step toward fault-tolerant quantum computing.
This structured co-design approach sets the foundation for the next sections, where we
dive deeper into device design parameters and compilation strategies. 

\subsubsection{Quantum Device Design}\label{subsec:device_design_parameters}
A comprehensive analysis of device-specific parameters is essential to understand their
impact on scalability and the feasibility of achieving practical utility in quantum computing.
Key factors include:
\begin{itemize}
    \item \textbf{Device Topology:} Physical arrangement and connectivity of qubits, which impacts quantum circuit mapping and execution efficiency. This includes the size and connectivity of the system back-end.
    \item \textbf{Native Gate Set:} Set of quantum gates natively supported by the respective
    hardware, which affects gate count and parallelisation.
    \item \textbf{Gate and Measurement Fidelity:} Probability of correctly applying
    quantum gates and accurately measuring states, which determine computational reliability.
    \item \textbf{Coherence Time:} Temporal stability of qubits before decoherence. Decoherence
    refers to the loss of a qubit's superposition state over time due to environmental
    interactions, causing the quantum state to collapse into a classical outcome~\cite{Schlosshauer_2019}.
    \item \textbf{Noise and Error Models:} The various sources of quantum errors that affect solution
    quality, including crosstalk, thermal relaxation, depolarisation noise, readout noise, and gate-based noise~\cite{Harper_2020}.
\end{itemize}
These parameters collectively define the operational envelope of a 
quantum processor. They not only determine which algorithms
can be realistically deployed, but also shape
compilation strategies, error mitigation techniques, and ultimately
the feasibility of achieving quantum advantage on current and 
near-term hardware~\cite{murali2019full}.

\subsubsection{Compiling Quantum Circuits to Quantum Hardware}
A systematic approach to quantum circuit compilation is essential to bridge the gap between abstract algorithms and hardware-specific execution.
To run a circuit on a given quantum processor, it must be transformed to
comply with hardware constraints, which vary across platforms.
These constraints, many of which are outlined in Section~\ref{subsec:device_design_parameters},
introduce complex challenges for efficient
execution~\cite{bandic2020structured}.
Addressing them requires a sequence of interdependent tasks, whose order
and implementation depend on the architecture and optimisation goals. The key tasks
are:
\begin{itemize}
    \item \textbf{Gate Decomposition:} The gates are transformed into the primitive
    gate set supported by the quantum processor~\cite{Rosa_2025}.
    \item \textbf{Operation Scheduling} Scheduling gates in parallelised manner to ensure minimal circuit depth and decoherence effects while respecting all the architecture-dependent shared control 
    constraints~\cite{guerreschi2018two}.
    \item \textbf{Qubit Allocation:} Assigns logical qubits to the physical 
    ones~\cite{qubitallocation}.
    \item \textbf{Qubit Routing:} Moves logical qubits into adjacency for \emph{two-qubit} interactions. For instance, by introducing SWAP gates~\cite{pant2019routing1,gyongyosi2020routing,doublystochasticrouting}.
    \item \textbf{Circuit Optimisation:} Performing tasks such as gate commutation or cancellation, transformation, with the aim of achieving the simplest form of the circuit~\cite{article_optimization_circ}.
    \end{itemize}
Our work provides a comprehensive framework for co-optimising
quantum software and informing hardware design. While it does
not directly optimise hardware components, it identifies
problem-dependent requirements and performance bottlenecks
that guide the specification of optimal hardware configurations.
This enables more informed algorithmic design choices and
supports the selection or development of hardware best suited
for targeted computational tasks.

\section{Methodology}\label{sec:methodology}
In this section, we outline the methodology employed to evaluate quantum circuit
performance under varying device and compilation conditions. We investigate the impact of different parameters as illustrated in Table~\ref{fig:parametersweep}.
Our experimental approach comprises two parts: one focusing on varying device parameters, while 
keeping compilation settings fixed, and the other examining the effects of varying compilation
conditions while maintaining fixed device parameters.
This dual approach allowed us to isolate and analyse the effect of device
variations independently from compilation variations, thereby providing a 
comprehensive understanding of these factors and their influence on 
quantum circuit performance under simulation.

\begin{table}[htbp]
\caption{Parameter ranges used in our experimental setup.}\label{fig:parametersweep}
  \begin{tabularx}{\linewidth}{XX}
 \toprule
  Parameter & Options \\\midrule
  \textbf{Device Design}\\\midrule
  Connectivity Density & [0.013895, 1.0]\\
  back-end & QASM Simulator (Qiskit)\\
  Native Gate Set & 'id', 'rz', 'sx', 'x', 'cx', 'swap', 'cz'\\
  Coupling Map & Heavy-Hex, Sycamore\\
  back-end Size (Heavy-Hex) & \(6\times 4\), \(6\times 5\), \(8\times 5\)\\
  back-end Size (Sycamore) & \(6\times 6\), \(11\times 11\), \(12\times 12\)\\
  Noise Model & Crosstalk, Thermal Relaxation, Depolarisation\\[1em]
  \midrule
  \textbf{Compiler Design}\\\midrule
  Optimisation Level & $0,1,2 $\\
  Layout Method & SABRE, Dense, Trivial\\
  Routing Technique & SABRE, Stochastic\\
  Additional Opt. Setups &  0-5\\
  Scheduling Method & ALAP\\
  \bottomrule
   \end{tabularx}
\end{table}

\subsection{Device Layer}\label{subsec:device_parameters}
In quantum computing, device parameters such as connectivity,
topology, and noise significantly play a critical role in
determining the fidelity and efficiency of quantum circuit execution.
These factors influence how circuits are mapped, routed, and scheduled,
often introducing overheads that significantly affect overall performance.
To ensure a comprehensive and systematic evaluation, we examine the impact of device
connectivity, topology configurations, back-end sizes and noise models. 
For experiments that involve varying device parameters, we employed fixed compiler
settings as follows: optimisation level $3$, qubit routing technique $\text{SABRE}$, 
layout method $\text{SABRE}$ and scheduling method \enquote{as late 
as possible}~(ALAP)~\cite{qiskit_transpiler}. 
A detailed analysis is presented in the subsequent section.

\subsubsection{Topology and Connectivity}\label{subsubsec:topologyandconnectivity}
\begin{figure}[htbp]
    \centering
    \begin{subfigure}[t]{0.47\textwidth}
        \centering
        \includegraphics[width=\linewidth]{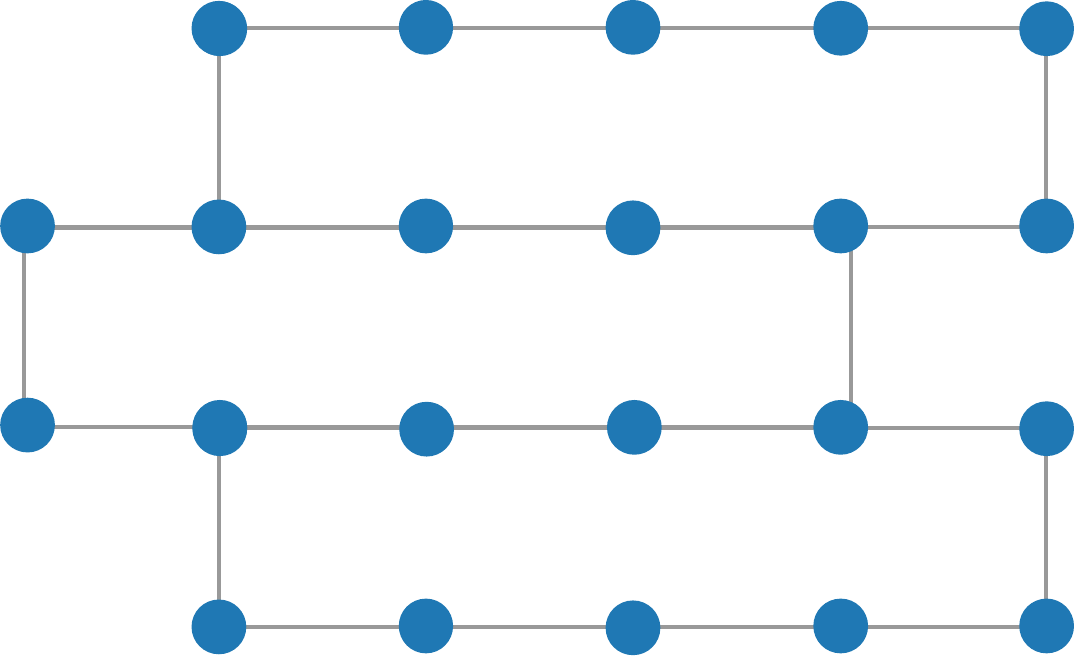}
        \caption{Heavy-hex topology}
        \label{fig:heavy-hex}
    \end{subfigure}
    \hfill
    \begin{subfigure}[t]{0.47\textwidth}
        \centering
        \includegraphics[width=\linewidth]{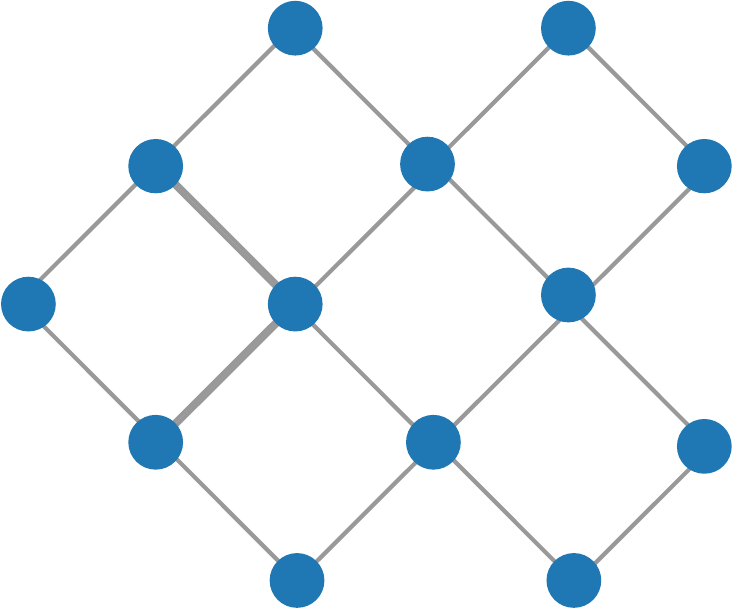}
        \caption{Grid-like topology of the Sycamore layout}
        \label{fig:sycamore}
    \end{subfigure}
    \caption{Comparison of hardware topologies.}
    \label{fig:topologies}
\end{figure}

The quantum circuits were compiled for two hardware architectures:
a heavy-hex and Sycamore, both illustrated in Figure~\ref{fig:topologies}.
Each architecture is characterised by a specific topology---the qubit
connectivity graph---and a corresponding physical layout,
which refers to the concrete arrangement of qubits on the chip.
The heavy-hex topology optimises connectivity while reducing crosstalk by limiting
qubit interactions to carefully placed neighbors~\cite{ibmq_heavy_hex},
whereas the grid like topology of the Sycamore chip uses a 2D grid where each qubit connects
to up to four neighbors, enabling efficient gate operations~\cite{arute2019quantum}.
To compare the topologies we introduce the following metric:
\begin{equation}
\text{Relative depth} = \frac{D_{\text{sycamore}} - D_{\text{heavy-hex}}}{D_{\text{heavy-hex}}},
\end{equation}
where \(D_{\text{sycamore}}\) and \(D_{\text{heavy-hex}}\) denote the circuit depths when
transpiled to the Sycamore and heavy-hex layouts.
This normalised measure captures the proportional increase or decrease in
circuit depth due to differences in 
topology and connectivity. It allows for comparisons by quantifying
how circuit execution complexity scales relative to a baseline, independent
of the absolute circuit size.
We study how varying the ratio of problem size to
back-end size affects circuit fidelity by uniformly scaling the physical layout.
By systematically adjusting the ratio of problem size to
physical qubit count, we analyse how the spatial embedding of a fixed logical 
circuit affects execution fidelity across different topologies
and device sizes.
When the number of qubits used to encode a problem is significantly
smaller than the size of the quantum processor, multiple logical-to-physical
qubit mappings become possible. This flexibility in placement
can influence circuit fidelity, as different mappings may
lead to variations in gate routing and noise accumulation. To explore
this effect, we vary the ratio of encoded problem size to back-end size
by uniformly scaling the physical layout.
back-end sizes are described in terms of their $2D$ as (\(n\times m\)), where \(n\) denotes
the number of rows, and \(m\) the number of columns.
Layouts implementing the heavy-hex topology range from configurations with
\(6\times 4\) ($143$ physical qubits) to \(8\times 6\) (297 physical qubits).
Similarly, Sycamore layouts
range from 36 qubits (\(6\times 6\)) to 144 qubits (\(12\times 12\)).

 The connectivity of the back-end is measured in terms of
a \emph{connectivity density}
\begin{equation}
c = \frac{N_C}{N_{C,\mathrm{max}}},
 \end{equation}
with $N_C$ denoting the total number of edges in the hardware
graph and $N_{C,\mathrm{max}}=N(N - 1)/2$ the maximal number of
edges for $N$ qubits. 
If two qubits are connected, they can physically interact and a
\emph{two-qubit} gate can be performed between them.
$c=1$ describes a device with all-to-all connectivity. 
In the experiments presented below, the connectivity density is increased
by randomly adding connections between qubit pairs that are not yet
connected, continuing until full connectivity is reached. The average number of nearest neighbours per qubit
grows linearly with the connectivity density.

\subsubsection{Noise} \label{subsubsec:noise}
In quantum computing, noise plays a critical role
in limiting circuit reliability and fidelity. We examine
common noise sources---crosstalk, thermal relaxation, and
depolarisation---and their impact on benchmark circuits.
Due to computational limitations, fidelities are approximated
analytically using a model that incorporates the number of gates,
gate fidelities, and circuit depth.

\paragraph{Crosstalk} refers to unintended interactions where
operations on one qubit affect nearby qubits, reducing fidelity.
It remains a hardware-specific challenge without a standardised model.
Inspired by prior studies on superconducting
qubits~\cite{paraskevopoulos2023spinq, ibm_crosstalk_10.1145/3373376.3378477},
we evaluate three crosstalk models — the shared qubit, the simultaneous execution,
and proximity based model — each capturing different aspects of crosstalk
behavior to assess their impact on circuit fidelity.

\textbf{Shared qubit} introduces crosstalk noise whenever 
two \emph{two-qubit} gates
share a common qubit, regardless of whether they occur in the same layer or at
different times in the circuit. 
This model is inspired by studies on superconducting qubits~\cite{ibm_crosstalk_10.1145/3373376.3378477, PRXQuantum.3.020301, PhysRevApplied.19.044078}. 
Prior research has shown that crosstalk effects during \emph{two-qubit} gate 
operations become significant when the gates share a neighboring pair.
A neighboring pair occurs when a qubit involved in one \emph{two-qubit}
gate is directly connected to a qubit involved in another \emph{two-qubit} gate.
Let qubits \(i\) and \(j\) participate in one \emph{two-qubit} gate operation, 
and qubits \(k\) and \(l\) in another. The crosstalk effect \(C(i, j, k, l)\) 
is defined as follows:
\[
C(i, j, k, l) =
\begin{cases} 
1 & \text{No connection between } i, j \text{ and } k, l, \\[0.5em]
C_{01,02} & \text{Connection between } i, j \text{ and } k, l.
\end{cases}
\] 
To quantify the effect of neighboring gates, the fidelity metric \(C_{01,02}\)
is defined as the harmonic mean of the fidelities of the overlapping \emph{two-qubit} gates:
\begin{equation}
    C_{01,02} = \left( \frac{2}{\frac{1}{F_{O_1}} + \frac{1}{F_{O_2}}} \right)^n
    \label{eq:crosstalk_metric}.
\end{equation}
Here:
\begin{itemize}
    \item \(F_{O_1}\) and \(F_{O_2}\): Fidelities of the two overlapping operations \(O_1\) and \(O_2\).
    \item \(n\): Degree of crosstalk amplification, proportional to the number \emph{two-qubit} gate operations executed.
\end{itemize}
The \textbf{simultaneous execution} model surfaces  when at least 
two \emph{two-qubit} gates are executed within the same circuit layer.
This model introduces crosstalk when 
\emph{two-qubit} gates are executed concurrently, regardless
of spatial separation, affecting neighboring qubits connected
to the active ones~\cite{PRXQuantum.3.020301}. For qubits \(i\) and \(j\) 
undergoing simultaneous \emph{two-qubit} operations at
time \(k\), the crosstalk effect \(C(i, j, k)\) is given by:
\[
C(i, j, k) =
\begin{cases} 
1 & \text{No operation on } i \text{ and } j \text{ at time } k, \\[0.5em]
C_{01,02} & i, j \text{ undergo ops. } O_1, O_2 \text{ at time } k.
\end{cases}
\]
The fidelity metric for simultaneous operations is similarly
to the shared qubit model expressed as:

\begin{equation}\label{Csimneigh_equation}
    C_{\text{neigh x sim}} = \prod_{(i,j) \in \mathcal{N}} F_{O_m}^n
    \left( \frac{2}{\frac{1}{F_{O_i}} + \frac{1}{F_{O_j}}} \right)^k,
\end{equation}
here:
\begin{itemize}
    \item \(F_{O_1}\) and \(F_{O_2}\): Fidelities of the two simultaneously executed \emph{two-qubit} gates.
    \item \(k\): Amplification factor for number of simultaneous \emph{two-qubit} operations.
    \item \(\mathcal{N}\): Set of all neighboring edges affected by the simultaneous \emph{two-qubit} gates.
    \item $F_{O_m}$ : Fidelity of the single-qubit operation on qubit .
    \item \(n\): Degree of influence, proportional to the number of shared qubits.
\end{itemize}
In addition to affecting the involved \emph{two-qubit} gates, both
models---\textbf{shared qubit} and \textbf{simultaneous execution}---also apply
a lower-intensity noise term to nearby \emph{single-qubit}
gates to reflect indirect interference.

\textbf{The proximity based} model considers crosstalk when \emph{two-qubit} gate operations are
executed within a predefined physical distance on the hardware coupling map.
Additionally, single-qubit gates performed on neighboring qubits are also affected by a
weaker noise penalty. Let qubits $i$ and $j$ participate in one \emph{two-qubit} gate operation,
and qubits $k$ and $l$ in another. If the Euclidean distance between any qubit from the
first operation and any qubit from the second operation is within a maximum
radius $r_{max} = 2$, crosstalk noise occurs. The crosstalk effect $C(i, j, k, l)$ is defined as follows:
\[
C(i, j, k, l) =
\begin{cases} 
1 & \text{if } d(i, j, k, l) > r_{\text{max}}, \\
C_{\text{prox}} & \text{if }d(i, j, k, l) \leq r_{\text{max}}.
\end{cases}
\]
Here, $d(i, j, k, l)$ is the Euclidean distance between any qubit
in the first gate operation and any qubit in the second gate operation.
The fidelity metric for proximity based crosstalk is given by:
\begin{equation}
C_{\text{total}} = C_{\text{prox}} \times C_{\text{neigh}},
\end{equation}
with
\begin{align}
    C_{\text{prox}} &= \left( \frac{2}{\frac{1}{F_{O_1}} + \frac{1}{F_{O_2}}} \right)^n,&
    C_{\text{neigh}} &= \prod_{m \in \mathcal{N}} F_{O_m}^n.
    \label{eq:c_neigh}
\end{align}
Here:
\begin{itemize}
\item $\mathcal{N}$: Set of all neighbouring qubits executing single-qubit gates.
\item $F_{O_m}$: Fidelity of the single-qubit operation on qubit .
\item $k$: Degree of influence, proportional to the number of affected neighbours.
\item \(F_{O_1}\) and \(F_{O_2}\): Fidelities of the two topologically close \emph{two-qubit} gates.
\item $n$: Amplification factor based on the proximity and number of affected qubits.
\end{itemize}

\paragraph{Other Noise Variants} 
We compare the effect of crosstalk to other prominent noise types---
thermal relaxation and depolarisation noise. By conducting
a systematic evaluation across these variants, we aim to 
quantify the relative significance of crosstalk noise in relation
to its counterparts. 
\textbf{Thermal relaxation} noise is a non-unital, irreversible
process describing the interaction between qubits and their 
environment as they evolve toward thermal equilibrium. A 
non-unital process does not preserve the identity operator, meaning
it drives the systems toward a specific state---in this case,
the ground state $\ket{0}$---rather than maintaining
a maximally mixed state. It comprises two primary mechanisms:
\begin{itemize}
    \item \textbf{Relaxation \(T_1\)}: The process by which
    a qubit exchanges energy with its environment, typically
    decaying from the excited state $\ket{1}$ to the ground 
    state $\ket{0}$. 
    \item \textbf{Dephasing \(T_2\)}: A process that
    leads to the decay of quantum coherence without necessarily
    changing the energy state of the qubit. 
\end{itemize}
These effects are characterised by timescales \(T_1^{(q)}\)
and \(T_2^{(q)}\) for each qubit \emph{q}, where typically \(T_2 \leq 2T_1\) ~\cite{Nielsen_Chuang_2010}.
Dephasing can occur both independently and in conjunction with
relaxation~\cite{PhysRevA.86.032324_Fowler}. 
To estimate fidelity loss due to thermal relaxation (amplitude
damping and dephasing) in a quantum circuit, we consider the total time each qubit is active during the 
execution of the circuit. Let \( t^{(q)} \) be the total accumulated gate duration for 
qubit \( q \). Each qubit is characterised by a \( T_1^{(q)} \) (energy relaxation) and \( T_2^{(q)} \) 
(dephasing) time constant. The fidelity due to thermal relaxation for qubit \( q \) is modeled as:
\begin{equation}
F^{(q)}(t^{(q)}) = e^{-t_q / T_1^{(q)}} \cdot e^{-t_q / T_\phi^{(q)}},
\end{equation}
where the pure dephasing time \( T_\phi^{(q)} \) is derived from:
\begin{equation}
\frac{1}{T_\phi^{(q)}} = \frac{1}{T_2^{(q)}} - \frac{1}{2 T_1^{(q)}}.
\end{equation}
The total circuit fidelity is estimated as the product of the 
fidelities over all qubits \( q \in Q \):
\begin{equation}
F_{\text{total}} = \prod_{q \in Q} F^{(q)}(t^{(q)}).
\end{equation}
This method provides a idealised, estimation of circuit fidelity
under thermal noise, assuming Markovian relaxation and no gate errors. 

\textbf{Depolarisation} noise affects quantum systems, by randomly replacing quantum states with the maximally
mixed state. This process results in a
complete loss of information about the original state, with the system transitioning
to a uniform probability distribution over all possible states. 
Depolarisation noise for a quantum gate is modelled using 
\begin{equation}
F_{\text{gate}} = F_{\text{initial}} \times \left( 1 - p_{\text{depolarisation}} \right),
\end{equation}
where $F_{\text{initial}}$ is the initial fidelity of the gate, and
$p_{\text{depolarisation}}$ is the depolarisation probability associated with the gate. 
The overall fidelity of a quantum circuit, incorporating all $N$ gates in 
the transpiled circuit, can be expressed as:
\begin{equation}
F_{\text{circuit}} = \prod_{i=1}^{N} F_{\text{gate},i},
\end{equation}
where $N$ is the total number of gates in the circuit~\cite{Cross_2016_depo,PhysRevA.104.062432_depo}.

\subsection{Compilation Layer}\label{subsec:Compilation_Methodology}
We evaluate the impact of compilation choices---such as optimisation level,
qubit mapping, and routing strategy---on circuit fidelity.
For the experiments exploring compilation variations, the fixed device 
configuration are set as follows: a coupling map with $128$ qubits
with connectivity densities set to $[0.013895,$ $0.03, $ $0.05, $ $0.1 ,$ 
$0.3, $ $0.5, $ $0.8]$. The basis gates are defined as $[$'x', 'y', 'z', 'rx'
'ry', 'rz', 'cx', 'cy'$]$.

\subsubsection{Metrics}
The circuit mapping performance metrics are defined as follows: 
Gate overhead measures the relative increase in the total number of quantum
gates after compilation:
\begin{equation}
    G_{\text{overhead}} = \frac{G_{\text{after}} - G_{\text{before}}}{G_{\text{before}}},
\end{equation}
where \( G_{\text{before}} \) and \( G_{\text{after}} \) denote the number of gates
before and after compilation. Similarly, depth overhead measures the
relative change in circuit depth with \( D_{\text{before}} \) and \( D_{\text{after}} \)
representing pre- and post- compilation:
\begin{equation}
    D_{\text{overhead}} = \frac{D_{\text{after}} - D_{\text{before}}}{D_{\text{before}}}.
\end{equation}
In the next expression, \( F_{\text{before}} \) is the fidelity before compilation and \( F_{\text{after}} \) is the fidelity after compilation.
\begin{equation}
    F_{\text{decrease}} = \frac{F_{\text{before}} - F_{\text{after}}}{F_{\text{before}}}.
\end{equation}
To measure the impact on solution quality,
we utilise the $\text{cost improvement}$ metric---as introduced in Arline Benchmarks~\cite{kharkov2022arlinebenchmarksautomatedbenchmarking}
---combining circuit depth, gate counts and gate fidelities.
The cost improvement is defined as the ratio between the initial
and final circuit costs. A higher ratio indicates the
optimisation of reduced errors.
This metric serves as the primary Figure of merit and builds upon the metric set previously 
introduced in~\cite{article_profiling_bandic}.
\begin{equation}
    C = \frac{C_{\text{in}}}{C_{\text{out}}},
    \label{eq:C_ratio}
\end{equation}
where:
\begin{align}
C_{\text{in}} &= - D_{\text{before}} \times \log K - 
N_{1q}^{\text{before}} \times \log F_{1q} 
- N_{2q}^{\text{before}} \times \log F_{2q},\label{eq:C_in}\\
C_{\text{out}} &= - D_{\text{after}} \times 
\log K - N_{1q}^{\text{after}} \times \log F_{1q} 
- N_{2q}^{\text{after}} \times \log F_{2q}.
\label{eq:C_out}
\end{align}

$F_{1q}$, $F_{2q}$ denote \emph{one-qubit} and \emph{two-qubit} gate fidelity (default: $0.9982$, $0.9765$), and 
$N_{1q}$ respectively $N_{2q}$ represent the number of single- and two-qubit gates before and after compilation.
The decoherence fidelity per depth unit, $K$ (default: $0.995$), 
models the loss in 
fidelity due to idle time and circuit depth, \ie, the longer
a qubit remains active within a deep circuit, the greater the
chance it suffers from decoherence effects. A value of $K = 0.995$ 
corresponds to a $0.5$\% fidelity loss per unit 
of depth~\cite{Nielsen_Chuang_2010}.

Although this work primarily focuses on
cost improvement metric as a comprehensive indicator for solution
quality, we have also evaluated the other three metrics: fidelity 
decrease, gate overhead, and depth overhead. These supplementary
results are included in the full set of experiments provided
in our reproduction package (see Section~\ref{sec:Introduction}).
Gate fidelities are derived from Starmon-5
a superconducting quantum processor based on circuit quantum electrodynamics~\cite{Starmon_5_Fact}. While our benchmark topologies are inspired
by IBM's heavy-hex and Google's Sycamore architectures, we chose
Starmon-5 gate fidelities as a more neutral and representative
reference point, avoiding direct bias toward a specific commercial 
platform. Furthermore, Starmon-5 exhibits performance
characteristics that fall within the typical range of superconducting
qubit platforms making it suitable for general benchmarking 
without favouring any particular topology or vendor.

\subsubsection{Optimisation Level}\label{subsub:opt_level}
Qiskit provides four optimisation levels (0–3) that progressively 
reduce circuit depth and gate count during transpilation, with higher
levels applying more aggressive transformations at the cost of longer
compilation time~\cite{optimization_level}.

\subsubsection{Layout Methods}\label{para:layout}
To map logical qubits to physical qubits on a quantum device, we
evaluate three Qiskit layout methods: Trivial, Dense, and SABRE~\cite{qiskit_transpiler}.
The \textit{Trivial} method maps logical to physical qubits in numerical
order without considering the hardware topology, making it computationally inexpensive
but potentially leading to many SWAP operations. 
The \textit{Dense} method aims to minimise SWAP operations by selecting
a subset of physical qubits that closely matches the logical qubit structure.
It analyses the device’s coupling map to identify a densely connected
group of qubits, reducing the distance between interacting qubits but
requiring more computational effort. The \textit{SABRE}
method uses a heuristic to iteratively refine the mapping as the circuit progresses, effectively minimising SWAP gates and thus circuit depth, especially in larger, more complex circuits.
\subsubsection{Qubit Routing Techniques}\label{para:routing}
Qiskit offers different routing techniques to insert SWAPs when logical
qubits are not physically adjacent~\cite{IBM_Routing}. We examine
\textit{Stochastic} and \textit{SABRE} routing: Stochastic uses
randomisation and heuristics to minimise circuit depth, while
SABRE dynamically adjusts qubit placement to reduce SWAP overhead during execution.
\subsubsection{Pass Manager Setups (Circuit Optimisation Passes)}\label{subsec:passmanager}
We investigate five pass manager setups, each applying progressively more complex
optimisation techniques to improve circuit fidelity.
\begin{itemize}
    \item \textbf{Setup 1}: Optimises \emph{single-qubit} gates and Clifford 
    operations by simplifying commutation relationships.
    \item \textbf{Setup 2}: Decomposes \emph{single-qubit} gates and
    cancels adjacent CNOT gates.
    \item \textbf{Setup 3}: Extends Setup $1$ by removing diagonal
    gates before measurements.
    \item \textbf{Setup 4}: Decomposes \emph{single-qubit} gates and applies
    commutative gate cancellation.
    \item \textbf{Setup 5}: Extends Setup $3$, applies the Hoare optimiser, 
    followed by (inverse) commutative gate cancellation.
\end{itemize}
While we utilise Qiskit as compilation, our
proposed DSE technique is compiler-agnostic and can be readily
applied within other quantum software platforms.

\subsection{Benchmarks}\label{subsec:benchmarks}
For our experiments, we selected an extensive set of quantum circuits that span a wide
range of computational paradigms and quantum algorithmic classes. The benchmark set 
includes circuits from well-known quantum algorithms and standard benchmarking
tools such as Quantum Volume, Shor's algorithm, 
Quantum Fourier Transform (QFT), and circuits generating 
Greenberger-Horne-Zeilinger (GHZ) states.
We include QFT and related circuits because
they exhibit well-understood structural and noise sensitivity
characteristics, making them suitable candidates for studying
residual noise effects and error mitigation techniques~\cite{Amy_2014}.
This is particularly important as the transition from NISQ
to FTQC is unlikely to be abrupt; intermediate regimes will
exhibit partial error correction and noise resilience,
for which techniques like those evaluated in this study are
highly relevant~\cite{Roffe_2019}.
Additionally,
we incorporated circuits for arithmetic operations, such as modular 
addition as well as structurally diverse randomly generated circuits
to capture non-algorithmic patterns and stress-test compiler
behaviour in less regular configurations. To further enhance
the diversity and structural variety of our benchmark set,
we ensured that the circuits represent distinct structural
categories based
on their circuit properties as defined in the paper 
by Bandic~\etal~\cite{article_profiling_bandic}. This clustering
is based on quantitative features rather than fixed
algorithm classes.
This approach guarantees that the
benchmarks are structurally distinct, enabling more comprehensive comparisons.
The benchmark set itself is derived from the qbench benchmark
set~\cite{bandic2023interaction, qbench}, which 
offers an extensive collection of quantum circuits, sourced
from various platforms and written in different programming
languages. To address contemporary challenges
we added error correction algorithms like Bosonic, Repetition, 
Shor, Steane and Surface Code. These circuits enable 
performance evaluation in the context of FTQC and
support the investigation of early-stage error correction
and mitigation strategies applicable to near-term devices.
The selected benchmarks enable a comprehensive evaluation of
gate-based quantum computing performance across diverse circuit
features and use cases, including optimisation, factoring and 
quantum simulation.  The full benchmark suite of 30 circuits 
along with information on each algorithm, its purpose, and 
implementation specifics are included in the reproduction 
package referenced in Section~\ref{sec:Introduction}.

\section{Results} 
We commence with discussing our results along the two parts
described in Section~\ref{sec:methodology}: Device and compilation 
characteristics.

\begin{figure}[htbp]
    \centering
    \includegraphics[width=\textwidth]{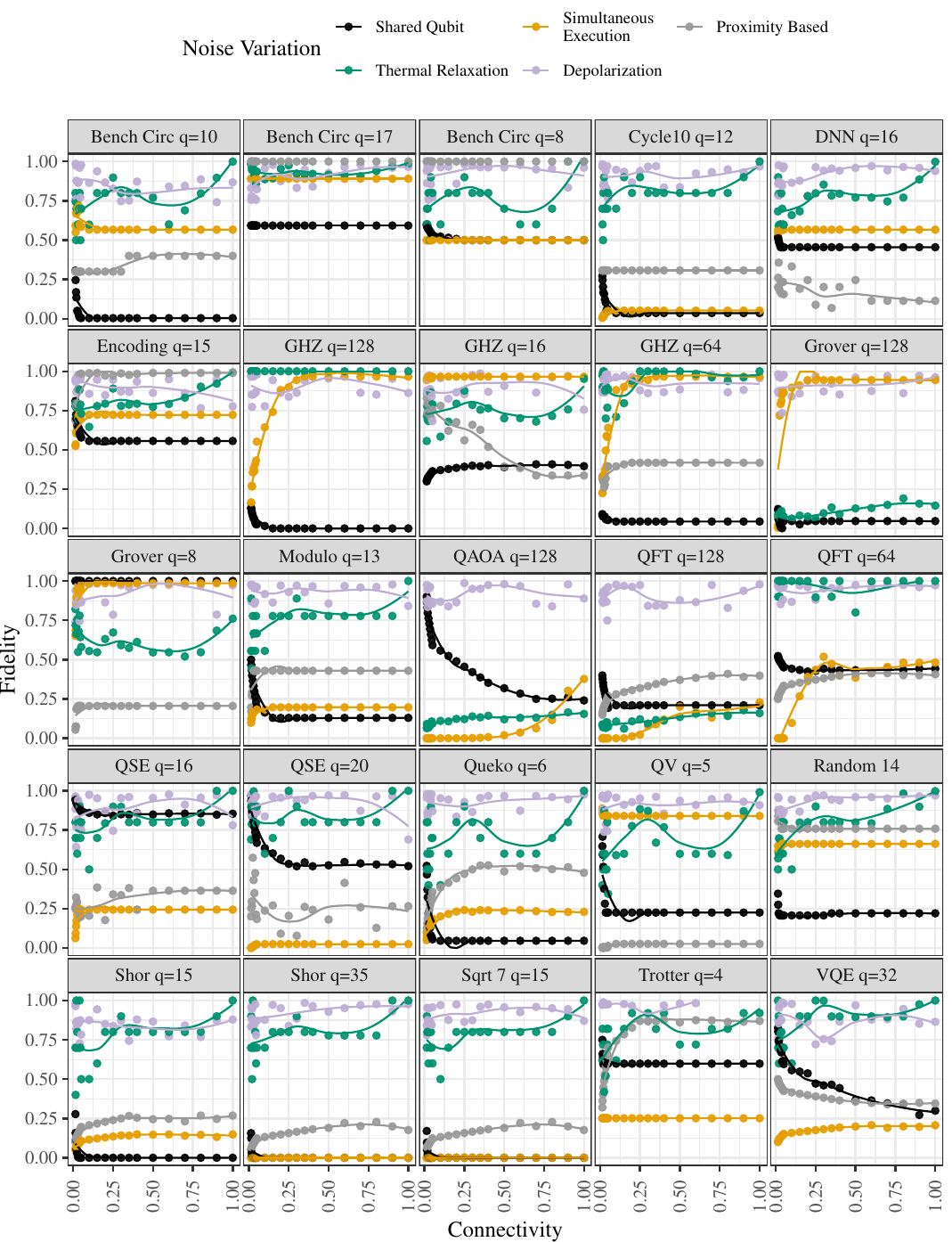} 
    \caption{Illustration of fidelity vs. connectivity across
    benchmarks as facets comparing three crosstalk models, thermal
    relaxation, and depolarisation noise for the heavy-hex back-end topology.}
    \label{fig:crosstalk_all_noises}
\end{figure}

\begin{figure}[htbp]
    \centering
    \includegraphics[width=\textwidth]{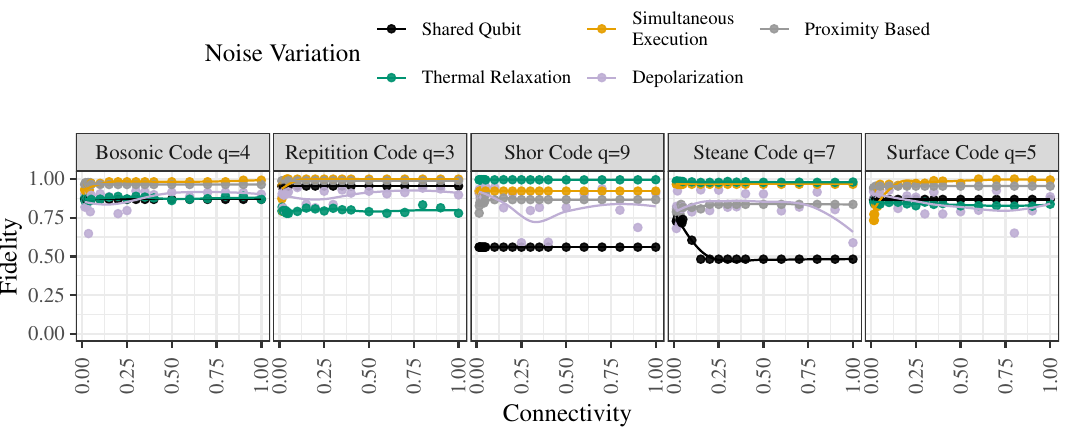} 
    \caption{Extension of Figure~\ref{fig:crosstalk_all_noises}
    with error correcting codes.}
    \label{fig:crosstalk_all_noises_ecc}
\end{figure}

\subsection{Results of the Device Setup and Parameter Sweep}
\begin{figure}[htbp]
    \centering
    \includegraphics[width=\textwidth]{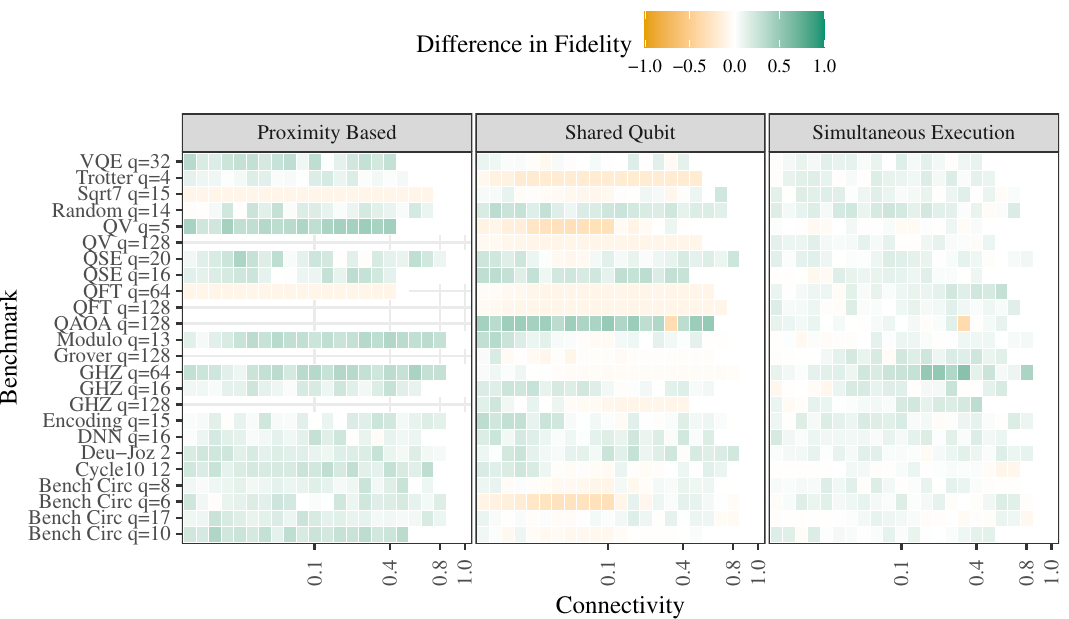} 
    \caption{Difference in fidelity between heavy-hex
    and the grid-like topology of the Sycamore chip across benchmarks and connectivity.
    }
    \label{fig:hhex_syc_depth}
\end{figure}

\begin{figure}[htbp]
    \includegraphics[width=\textwidth]{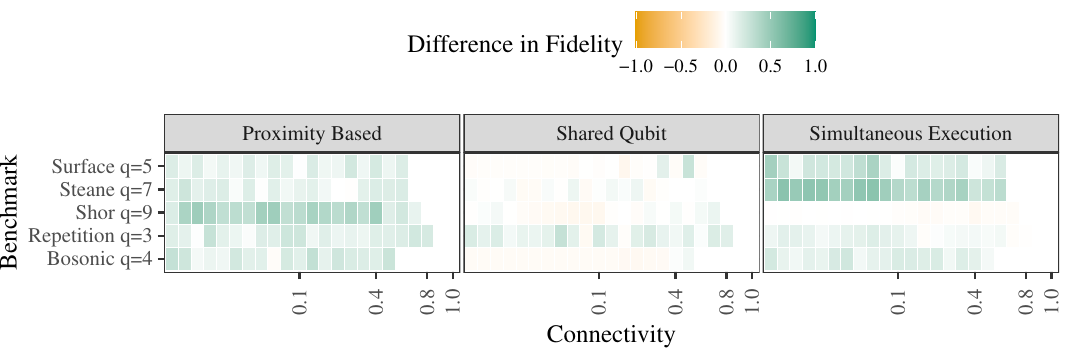} 
    \caption{Extension of Figure~\ref{fig:hhex_syc_depth}
    with error correcting codes.
    }
    \label{fig:hhex_syc_depth_ecc}
\end{figure}

To isolate the impact of each noise model, we evaluate circuit fidelity
and circuit depth under five distinct noise models, considering 
them individually and independently in an otherwise noiseless setting.
The three crosstalk-related models include
\textbf{shared qubit}, \textbf{simultaneous execution}, and
\textbf{proximity based} (see Section~\ref{subsubsec:noise}),
each representing plausible physical behaviours in 
multi-qubit gate execution systems. In addition, we include two
widely studied standard noise models: \textbf{thermal relaxation}
and \textbf{depolarisation}. 
Figure~\ref{fig:crosstalk_all_noises} compares the effects of all five
noise models across varying connectivity densities for a representative set of
benchmarks (see Section~\ref{subsec:benchmarks}).
Note that the fidelities are model-estimated and may introduce bias or
approximation artifacts, particularly in regimes with correlated errors.
We observe that devices with higher connectivity density (right-hand side of the x-axis)
are generally more resilient to the
simultaneous execution model, as parallel gate execution causes less disruption 
when the qubit layout is less constrained. In contrast, the shared qubit model
highlights a trade-off where more densely connected devices can amplify
interference and common qubits. The proximity based model
performs more consistently across connectivity densities, indicating
its relative independence from overall connectivity density.
No single crosstalk model dominates across all benchmarks. Which model
performs best is highly benchmark-specific, reinforcing the need for tailored
noise mitigation. That said, the shared qubit model nearly consistently results
in the most severe degradation, highlighting it as a key target
for error mitigation in quantum architectures with slightly higher connectivity density.
Beyond crosstalk, quantum circuits are also affected by thermal relaxation
and depolarisation noise. While thermal relaxation and depolarisation are 
generally less harmful to fidelity than crosstalk---particularly compared to the 
shared qubit model---exceptions do exist. Notably, in large-qubit
benchmarks such as \emph{QAOA q=128}, \emph{Grover q=128}, and \emph{QFT q=128},
thermal relaxation causes greater fidelity loss than any of the other
noise variants. Error-correcting codes are essential for achieving
fault-tolerant quantum computing (FTQC). To evaluate their resilience
to crosstalk, we analyse the fidelity of well-known error-correcting codes, as shown in
Figure~\ref{fig:crosstalk_all_noises_ecc}. Proximity based and simultaneous execution models
maintain stable fidelities for all codes when connectivity exceeds $0.1$. However,
the shared qubit model shows a fidelity drop with increasing connectivity, particularly impacting
steane code. Shor code performs significantly worse under the shared qubit model than
the other two models. These results highlight the need for tailored error-correcting strategies
in FTQC, that consider both noise models and hardware topology.
Effectively mitigating crosstalk remains a complex challenge.
For example, Qiskit's now removed \texttt{CrosstalkAdaptiveSchedule}~\cite{crosstalk_adaptive_schedule}
aimed to reduce crosstalk by locally adapting gate scheduling.
However, it was found that such local optimisations can have
unpredictable global effects on circuit performance by increasing
exposure to other noise sources. This illustrates
the difficulty of balancing different noise mechanisms and 
underscores the need for community-wide standards to define how
crosstalk is modeled, measured, and mitigated.
Nevertheless, new techniques
like twirling~\cite{Twirling} and dynamical decoupling~\cite{article_dd} are emerging as promising techniques to reduce crosstalk error. Tunable couplers~\cite{TunableCouplers}
have also been proposed as a hardware-level solution to suppress crosstalk by
enabling dynamic control over qubit-qubit interactions.
To further investigate the spatial aspects of crosstalk, Figure~\ref{fig:hhex_syc_depth}
and~\ref{fig:hhex_syc_depth_ecc} compare
the heavy-hex and Sycamore layouts, each consisting of 143 and 144 qubits, respectively.
Positive values indicate better performance on heavy-hex, while negative values
favour Sycamore. Overall, Sycamore shows lower robustness to crosstalk across most
models and circuits, with few exceptions. This highlights the importance of
back-end-specific co-design. This is particularly evident when considering that
Qiskit's transpiler has been shown to perform especially well on the 
heavy-hex architecture~\cite{nation2025benchmarkingperformancequantumcomputing,ibmq_heavy_hex}.
However, the advantages of adopting a different topology diminish around a connectivity
density of $0.8$. For certain smaller benchmarks, this effect appears even earlier. 

To assess further architectural impacts on performance,
we examine how different back-end sizes influence fidelity and circuit depth.
\begin{figure}[htbp]
    \centering
    \includegraphics[width=\textwidth]{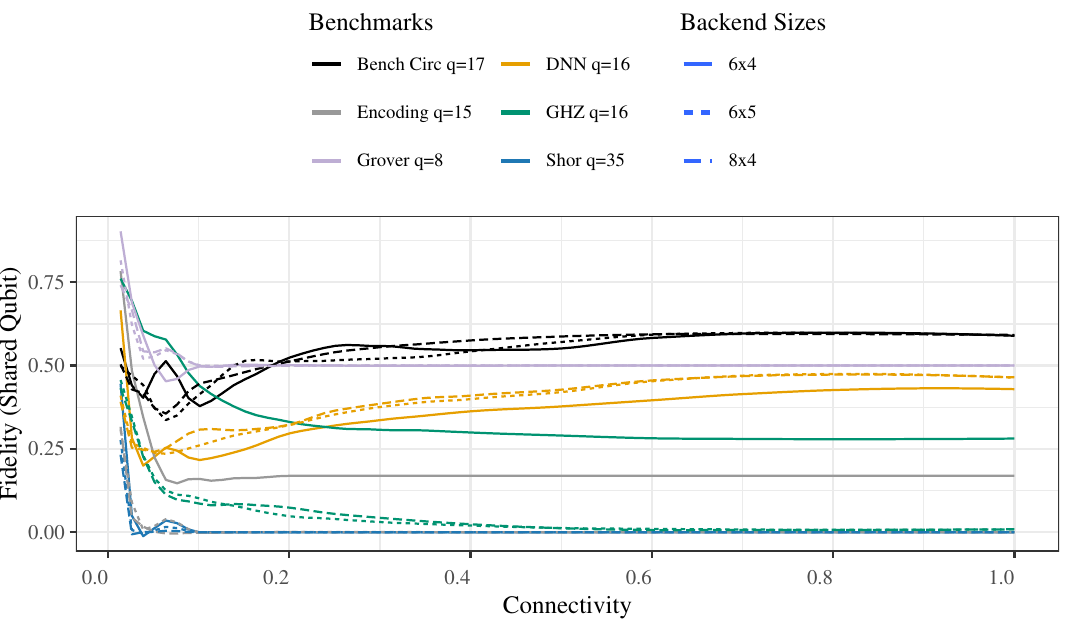} 
    \caption{Impact of back-end size fidelity as
    a function of connectivity across benchmarks for
    crosstalk version shared qubit.}
    \label{fig:back-endsize_cx}
\end{figure}
As seen in~\ref{fig:back-endsize_cx}, different
back-end sizes across different optimisation levels ($0-3$) have little 
to no consistent effect on
fidelity across benchmarks. In certain cases,
larger back-ends even exhibit a lower fidelity,
suggesting that scaling alone is insufficient---
if not a disadvantage---for mitigating crosstalk-induced
loss.

\begin{figure}[htbp]
    \centering
    \includegraphics[width=\textwidth]{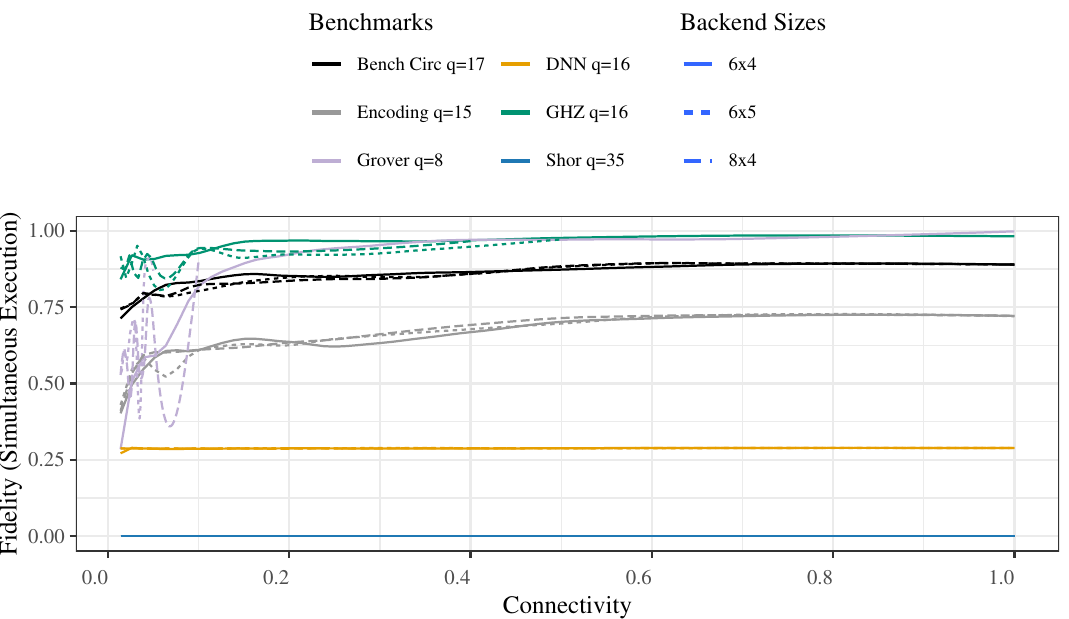} 
    \caption{Impact of back-end size fidelity as
    a function of connectivity across benchmarks for
    crosstalk version simultaneous execution.}
    \label{fig:back-endsize_ncx}
\end{figure}

\begin{figure}[htbp]
    \centering
    \includegraphics[width=\textwidth]{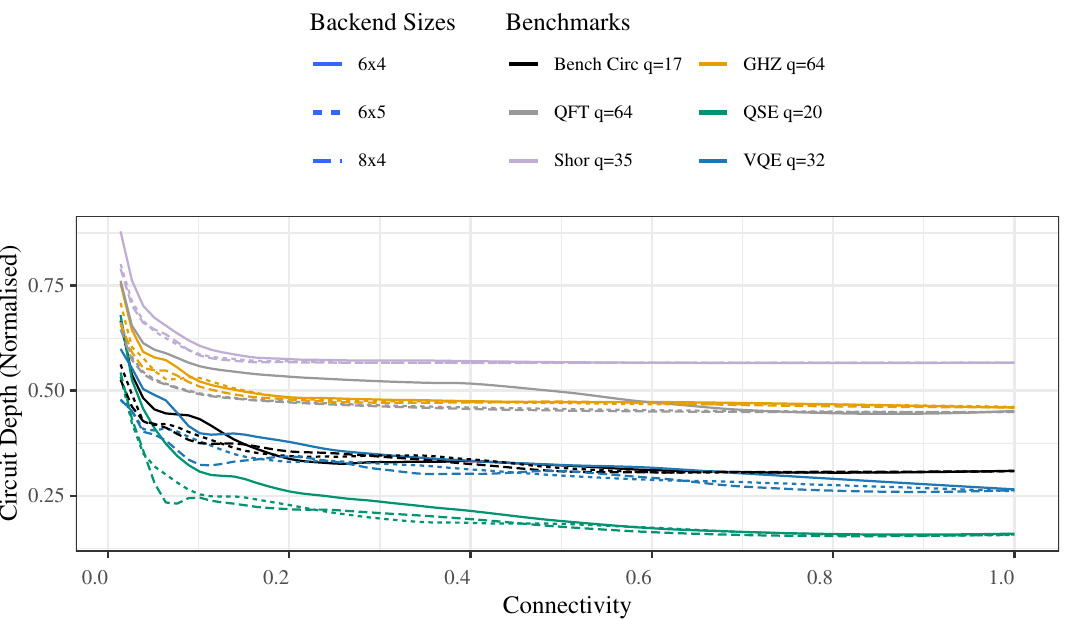} 
    \caption{Impact of back-end size on normalised circuit depth as
    a function of connectivity across benchmarks.}
    \label{fig:depth_hhex}
\end{figure}

The results in Figure~\ref{fig:back-endsize_ncx} reinforce the observation
that back-end sizes exerts minimal influence on 
fidelity. Under the simultaneous execution crosstalk model,
increasing the back-end size does not yield noticeable 
improvements in fidelity.
While some minor fluctuations are visible, the overall
trend indicates that fidelities remain largely
unaffected by back-end size scaling.
Figure~\ref{fig:depth_hhex} indicates that the depth of most circuits converges
near a connectivity of 0.3, regardless of back-end size. This convergence point is also
consistent with findings in optimisation-focused problems, as
reported by Safi~\etal~\cite{Safi_10234303}.
While most larger benchmarks also tend to stabilise around
this connectivity threshold, some continue to exhibit
slight reductions in circuit
depth beyond a connectivity of $0.3$. However, these gains
are marginal and often come at the cost of increased
compilation time. In general, back-end size has a
limited impact on depth and fidelity. Selecting a back-end
size that matches the scale of the target problem helps strike a
balance between computational efficiency and
resource allocation.

Overall, the results presented in this section indicate that higher 
connectivity density improves resilience to crosstalk noise across most models.
Nonetheless, fidelity in the shared qubit model can degrade as connectivity
increases. The grid-like topology of the Sycamore chip
exhibits lower robustness compared to its counterpart, and both fidelity and circuit depth
converge independent of back-end size.

\subsection{Results of the Compilation Parameter Sweep}
Having examined device-level parameters, we now focus on 
compilation parameters. We sweep the full set of benchmarks 
across all combinations as defined in Section~\ref{sec:methodology}. 

\begin{figure}[htbp]
    \centering
    \includegraphics[width=\textwidth]{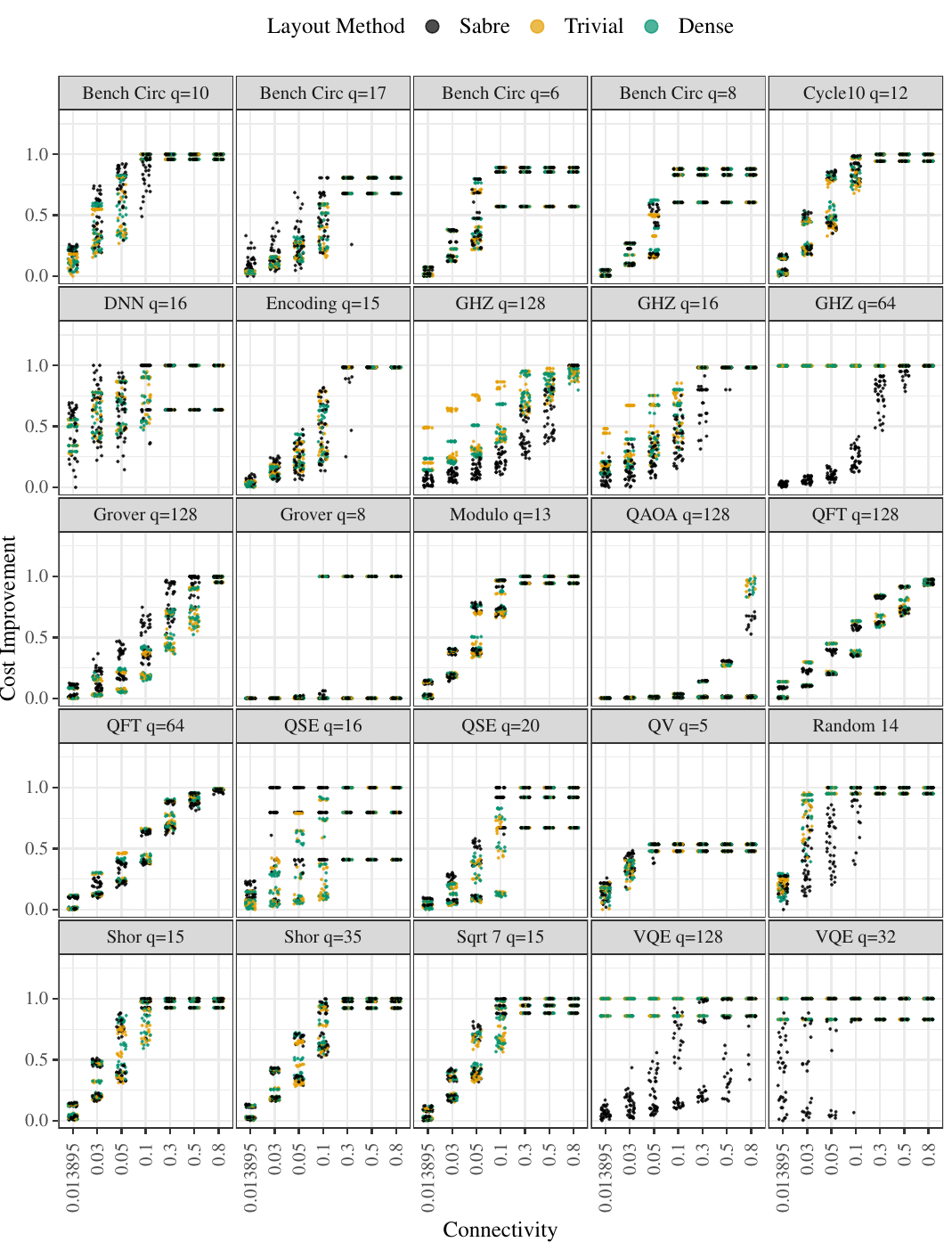} 
    \caption{Cost improvement (see Equations \eqref{eq:C_ratio} \eqref{eq:C_in}, \eqref{eq:C_out}) vs. connectivity across
    benchmarks for three layout methods.}
    \label{fig:layout_method}
\end{figure}

As shown in Figure~\ref{fig:layout_method}, we evaluate
each benchmark using the cost improvement metric introduced in section methodology (see Equations~\eqref{eq:C_ratio}, \eqref{eq:C_in}, and~\eqref{eq:C_out}), where
higher values indicate better performance. 
Across most 
benchmarks, the SABRE and Dense layout provide the best improvements,
though Trivial performs well for specific cases like GHZ states.
The Dense layout shows variable results, often outperforming Trivial
and sometimes matching SABRE. While SABRE yields high-quality mappings
it also has the longest compilation time.
Moreover, its performance deteriorates for 
variational algorithms such as VQE, where repeated circuit
execution favours stable and noise-aware mappings
~\cite{10.1145/3297858.3304007}.
Therefore,
its use should be carefully considered in light
of both time constraints and the specific characteristics of the problem.

\begin{figure}[htbp]
    \centering
    \includegraphics[width=\textwidth]{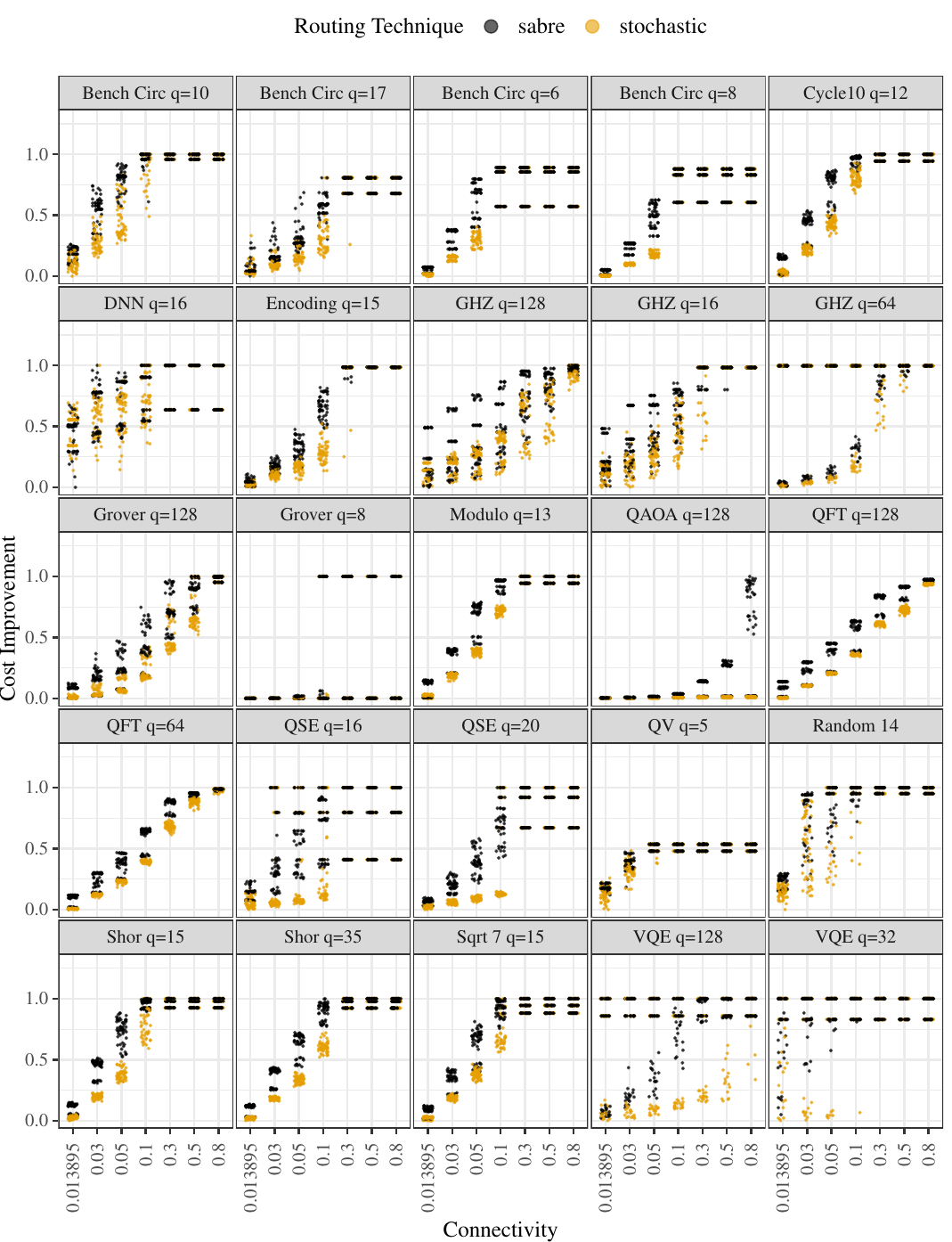} 
    \caption{Illustration of cost improvement as defined in Equations \eqref{eq:C_ratio} \eqref{eq:C_in}, \eqref{eq:C_out} vs. connectivity across
    benchmarks as facets comparing two routing techniques.}
    \label{fig:routing_technique}
\end{figure} 

Regarding qubit routing technique (see Figure~\ref{fig:routing_technique}), 
the SABRE router consistently outperforms the Stochastic approach, especially for large and structurally complex circuits such as QFT.
However, this advantage has trade-off's. The SABRE method typically 
requires longer compilation time (\ref{para:routing}), and for 
smaller or highly regular circuits, the improvement over Stochastic routing does 
not justify the additional effort.
However unlike the layout methods, where the optimal choice is problem dependent,
the comparison of qubit routing techniques reveals a clear trend.

\begin{figure}[htbp]
    \centering
    \includegraphics[width=\textwidth]{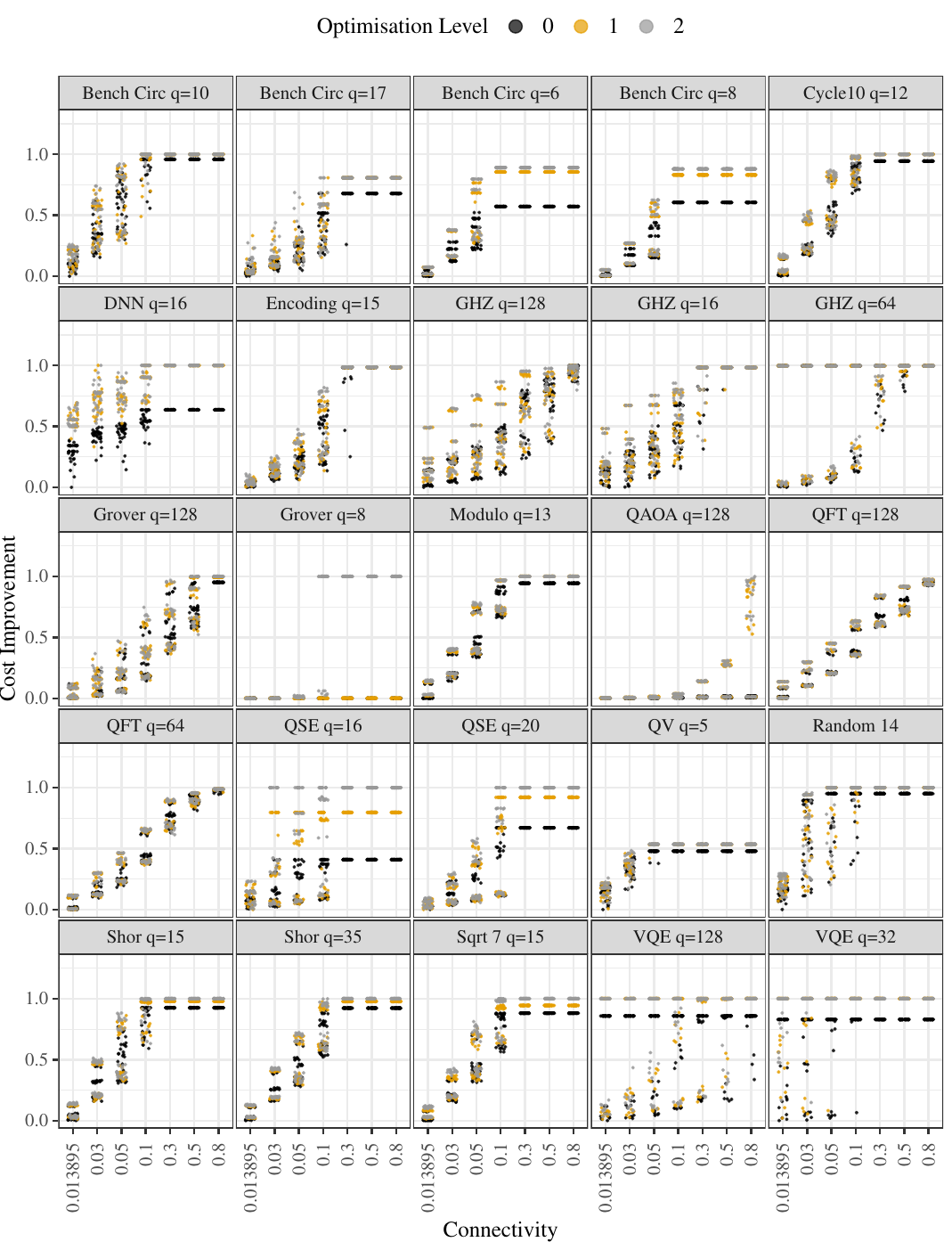} 
    \caption{Illustration of cost improvement as defined in Equations \eqref{eq:C_ratio} \eqref{eq:C_in}, \eqref{eq:C_out} vs. connectivity across
    benchmarks as facets comparing three optimisation levels.}
    \label{fig:optimisation_level}
\end{figure}
Figure~\ref{fig:optimisation_level} illustrates the impact of various
optimisation levels on cost improvement. As optimisation level $2$ rarely offers significant advantages
over level $1$, the additional compile-time overhead and
increased variability make optimisation level $1$ the preferred choice across most benchmarks.

\begin{figure}[htbp]
    \centering
    \includegraphics[width=\textwidth]{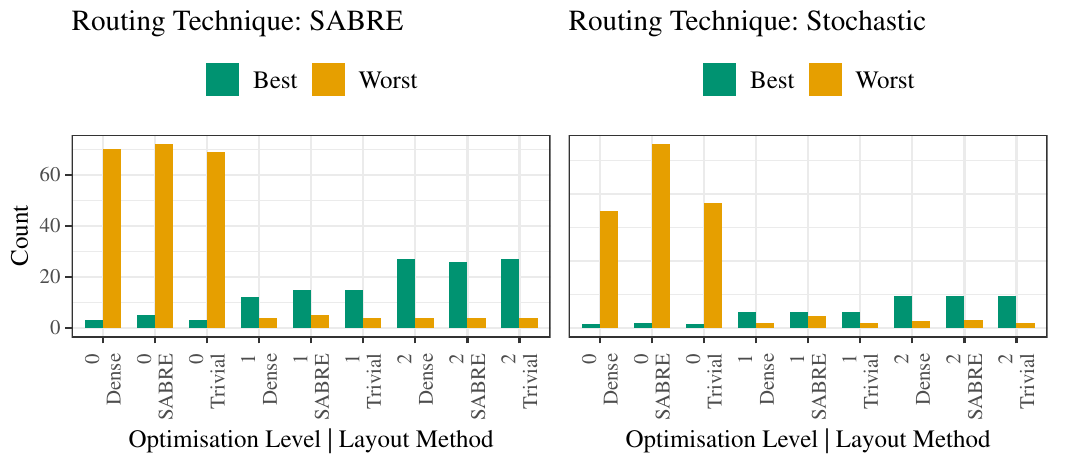} 
    \caption{Frequency of best and worst-performing
    combinations, based on cost improvement, evaluated across qubit routing technique, layout method, and optimisation level over
    all benchmarks and connectivity densities}
    \label{fig:best_worst_cost_comparison}
\end{figure}

\begin{figure}[htbp]
    \centering
    \includegraphics[width=\textwidth]{
    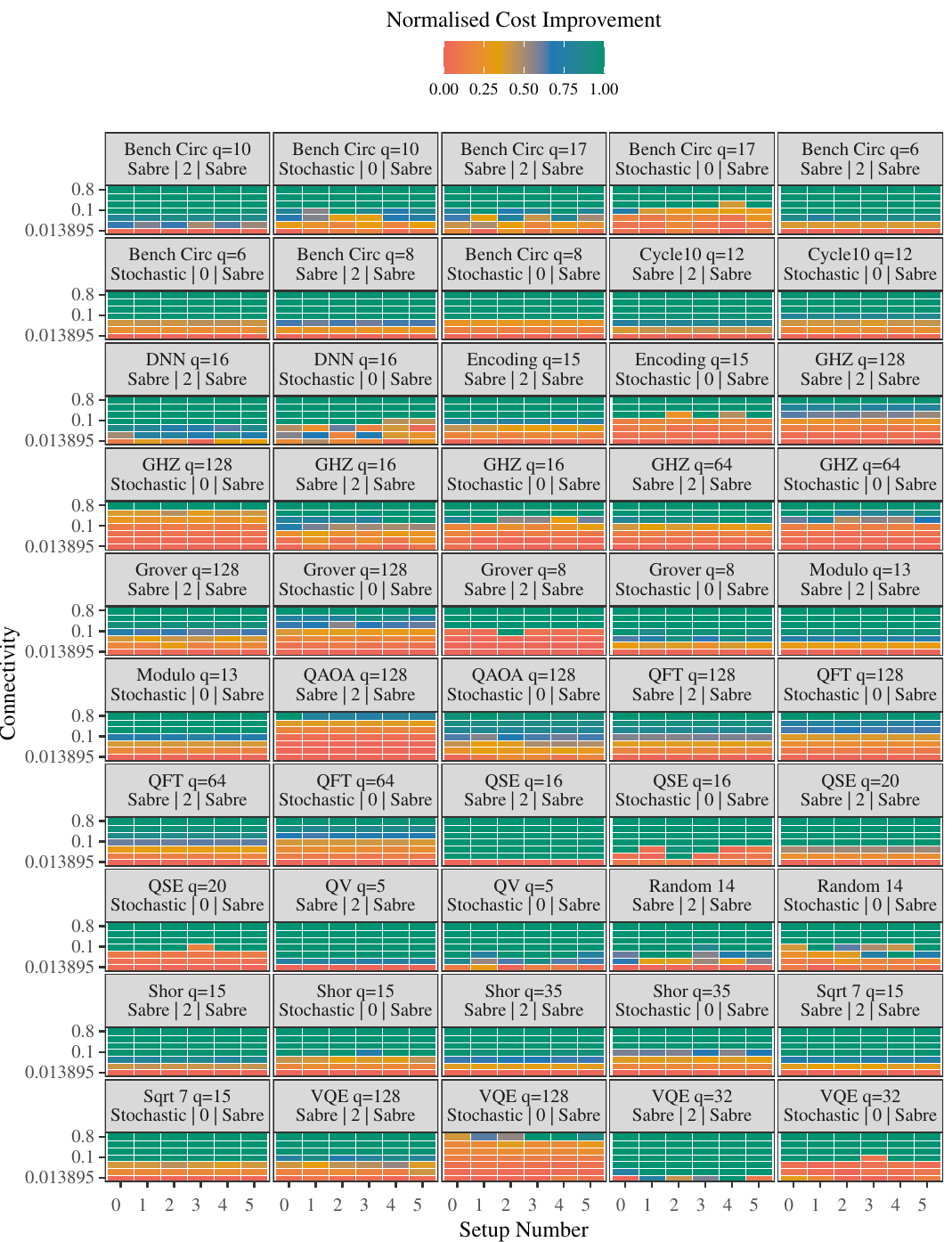} 
    \caption{Cost improvement (see Equations~\ref{eq:C_ratio}~\ref{eq:C_in}, \ref{eq:C_out}) vs. connectivity across benchmarks and 
    compilation setups. Improvements are measured relative to the
    baseline configuration with no additional improvements.}
    \label{fig:Sabre_2_Sabre}
\end{figure}

Although analysing layout method, qubit routing technique, and optimisation
level separately is useful, their effects are often interdependent.
Certain combinations can perform notably better or worse due 
to synergies between them.
Figure~\ref{fig:best_worst_cost_comparison} presents 
the frequency of best and worst-performing configurations
across all benchmarks. The most successful combinations
typically use SABRE as a qubit routing technique with optimisation level $2$,
with SABRE\textbar{}2\textbar{}SABRE performing best overall, followed by 
SABRE\textbar{}2\textbar{}Trivial and SABRE\textbar{}2\textbar{}Dense.
The worst results often come from Stochastic\textbar{}0\textbar{}SABRE, further validating 
that strong components alone are not enough---effective performance
requires aligned configurations across layout method, qubit routing technique, and optimisation level.
Notably, even SABRE\textbar{}0\textbar{}SABRE performs poorly, underlining
the significant role of the optimisation level.
Figure~\ref{fig:Sabre_2_Sabre} compares the performance
of the best and worst-performing initial configurations 
according to the cost improvement
metric across our benchmark algorithms and different connectivity
densities. The y-axis indicates the connectivity density
of the hardware topology, while the x-axis represents
successive passes introducing additional circuit 
optimisations (as discussed in
Section~\ref{subsec:passmanager}) aimed at reducing
depth and gate overhead, as well as improving fidelity.
Configurations that already incorporate an
effective combination of layout method, optimisation level,
and qubit routing technique generally do not benefit from 
additional circuit transformation passes. Even the 
worst-performing initial configurations tend to benefit 
minimally from additional passes, showing that
further complexity does not necessarily yield in
proportional gains.
Ultimately, the most significant factor
in improving circuit performance is connectivity: the more complex
the circuit, the more essential high connectivity becomes.
Generally, in combination SABRE consistently delivers the best layout
and routing performance but has higher compile times. Thus
optimisation level 1 offers a good compromise.

\clearpage  
\section{Conclusion \& Outlook}
Our study highlights the importance of a holistic approach to optimising
quantum circuit performance, emphasising the interplay between hardware
characteristics and compilation strategies. The results demonstrate that
both device attributes and compilation choices impact circuit performance.
We observe that different crosstalk models affect fidelity in distinct ways,
underlining the importance of tailoring error mitigation strategies to the
specific noise characteristics of the device. Notably, the shared qubit model
consistently exhibits the most detrimental impact on fidelity, while the
simultaneous execution model shows a more stable behavior, especially when
connectivity exceeds $0.3$. However, full connectivity is not required
as fidelity converges much sooner.
Additionally, back-end-specific optimisations play a crucial role.
Heavy-hex topology consistently shows better crosstalk resilience compared
to the grid-like topology of the Sycamore chip. In contrast larger back-end sizes have a
negligible impact on circuit fidelity.
Among compilation strategies, SABRE-based routing and layout methods with 
moderate optimisation levels yield the best compromise between
performance and compile time.
Notably, increasing system complexity through excessive optimisation 
does not lead to proportional improvements in performance, 
as shown in Figure~\ref{fig:Sabre_2_Sabre}. This suggests diminishing
returns beyond setting good parameters for qubit routing, layout method
and optimisation level.
Importantly, optimal performance
emerges from aligning all parameter options.
Our findings support the need for end-to-end co-design:
quantum system performance is maximised when noise-aware
hardware selection, connectivity, and compilation strategies 
are considered jointly.
Key takeaways include:
\begin{itemize}
    \item Connectivity density is the dominant factor influencing
    both fidelity and circuit depth.
    \item The shared qubit crosstalk model presents the greatest
    fidelity challenge and should be prioritised in mitigation 
    strategies.
    \item back-end size scaling offers limited to no benefit, suggesting
    resource-aware deployment is preferable.
    \item Among compilation parameters, SABRE routing combined with optimisation
    level 1 provides the best cost-performance balance.
    \item Adding more circuit transformations beyond already aligned
    configurations yields marginal or no gains, reinforcing the value of 
    strategic simplicity.
\end{itemize}
Building on these insights, hardware developers
should address some issues at the design level. 
It is beneficial to establish a standardised 
definition of crosstalk to ensure more consistent benchmarking.
Adaptive compilation strategies that respond to 
both circuit structure and device properties hold promise
for scalable quantum computing. Ultimately, aligning hardware
innovations with software design will be the key to achieving
the full potential of quantum technologies.
In future work, it would be valuable to investigate the
trade-offs between circuit performance and the additional compilation
time by more advanced circuit improvements.

\section*{Acknowledgments}
MB thanks H.~van Someren, P.~le Henaff, R.~Turrado Camblor and A.~Hesam for insightful discussions.
SF and MB would like to acknowledge funding from Intel Corporation. CGA is thankful for support from the QuantERA grant EQUIP with the grant number PCI2022-133004, funded by Agencia Estatal de Investigación, Ministerio de Ciencia e Innovación, Gobierno de España, MCIN/AEI/10.13039/501100011033, and by the European Union NextGenerationEU/PRTR. 
Also from the Ministry for Digital Transformation and of Civil Service of the Spanish Government through the QUANTUM ENIA project call - Quantum Spain project, and by the European Union through the Recovery, Transformation and Resilience Plan - NextGenerationEU within the framework of the Digital Spain 2026 Agenda.
This work was supported by the German Federal Ministry of Education and
Research (BMBF), funding program ‘quantum technologies---from basic research to market’, grant numbers 13N16092 and 13N16093. 
This work used the Dutch national e-infrastructure with the support of the SURF Cooperative using grant no. EINF-5822.
WM acknowledges support by the High-Tech Agenda of the Free State of Bavaria.

\bibliography{sn-bibliography}

\end{document}